\documentclass[aps,prb,twocolumn,showpacs,superscriptaddress]{revtex4-1}
\usepackage{graphicx}
\usepackage{dcolumn}
\usepackage{bm}
\usepackage{hyperref}
\usepackage[utf8]{inputenc}
\usepackage[english]{babel}
\usepackage{units}
\usepackage{mathtools}
\hypersetup{colorlinks=true,citecolor={blue},linkcolor={blue},urlcolor={blue}}
\usepackage{amsmath}
\usepackage{empheq}
\usepackage{mathrsfs}
\usepackage{amssymb}
\usepackage{soul}
\setstcolor{red}

\begin{document}

\title{Optical orientation, polarization pinning, and depolarization dynamics in optically confined polariton condensates}

\author{I. Gnusov}
\email[I. Gnusov ]{Ivan.Gnusov@skoltech.ru}
\address{Skolkovo Institue of Science and Technology, Novaya St. 100, Skolkovo 143025, Russian Federation.}

\author{H. Sigurdsson}
\address{Skolkovo Institue of Science and Technology, Novaya St. 100, Skolkovo 143025, Russian Federation.}
\address{School of Physics and Astronomy, University of Southampton,  Southampton, SO17 1BJ, UK.}

\author{S. Baryshev}
\address{Skolkovo Institue of Science and Technology, Novaya St. 100, Skolkovo 143025, Russian Federation.}

\author{T. Ermatov}
\address{Skolkovo Institue of Science and Technology, Novaya St. 100, Skolkovo 143025, Russian Federation.}

\author{A. Askitopoulos}
\address{Skolkovo Institue of Science and Technology, Novaya St. 100, Skolkovo 143025, Russian Federation.}

\author{P. G. Lagoudakis}
\email[P.G.Lagoudakis ]{P.Lagoudakis@skoltech.ru}
\address{Skolkovo Institue of Science and Technology, Novaya St. 100, Skolkovo 143025, Russian Federation.}
\address{School of Physics and Astronomy, University of Southampton,  Southampton, SO17 1BJ, UK.}

\begin{abstract}
We investigate the optical orientation, polarization pinning, and depolarization of optically confined semiconductor exciton-polariton condensates. We perform a complete mapping of the condensate polarization as a function of incident nonresonant excitation polarization and power. We utilize a ring-shaped excitation pattern to generate an exciton-induced potential that spatially confines polariton condensates into a single mode. We observe that formation of circular polarization in the condensate persists even for a weakly cocircularly polarized pump. By varying the excitation ring diameter we realize a transition from the condensate polarization being pinned along the coordinate-dependent cavity-strain axes, to a regime of zero degree of condensate polarization. Analysis through the driven-dissipative stochastic Gross-Pitaevskii equation reveals that this depolarization stems from a competition between sample induced in-plane polarization splitting and the condensate-reservoir overlap. An increase in the role of the latter results in weakening of the condensate fixed-point phase space attractors, and enhanced random phase space walk and appearance of limit cycle trajectories, reducing the degree of time-integrated polarization.
\end{abstract}
\maketitle

\section{Introduction}
Exciton-polaritons (from here on {\it polaritons}) are quasiparticles that arise from the strong coupling between exciton and photon modes in semiconductor microcavities~\cite{kavokin_microcavities_2007}. Due to their composite nature they are characterized by a low effective mass, short lifetime, and strong nonlinearity which has made them a popular candidate to explore novel nonlinear physics. Despite their short lifetime they can be sustained in stable form through various excitation schemes with their particle state information encoded in the emitted cavity light, including spin structure. Of interest, when polaritons condense~\cite{kasprzak_bose-einstein_2006, balili_bose-einstein_2007} one gains access to ultrafast nonlinear spin dynamics with possible applications in spinoptronic devices~\cite{pol_review,Liew_PhysE2011}.

Because of their two-level spin structure, polaritons offer an exotic platform to study various non-Hermitian spin physics. Their sensitivity to external magnetic fields~\cite{Larionov_PRL2010, Caputo_CommPhy2019}, cavity mirror birefringence~\cite{wang_polarization-coupled_2015}, along with a unique spin-orbit coupling mechanism known as the optical spin Hall effect~\cite{leyder_observation_2007,kammann_nonlinear_2012} has paved the way for the realization of polaritonic Chern insulators~\cite{klembt_exciton-polariton_2018,Nalitov_PRL2015, Bardyn_PRB2015}, polarized solitons~\cite{Sich_ACSPho2018} and half skyrmions~\cite{cilibrizzi_PRB2016}, spin switches~\cite{Amo_NatPho2010,askitopoulos_nonresonant_2016, Dreismann_NatMat2016}, spontaneous lattice ordered polarization~\cite{ohadi_spin_2017}, spin-selective filters~\cite{Gao_APL2015}, spin bistability~\cite{pickup_optical_2018, Ohadi19}, spin multistabilty~\cite{Paraiso_Nature2010}, spin valves~\cite{askitopoulos_all-optical_2018}, and measuring the quantum geometric tensor~\cite{Gianfrate_Nature2020}. 

Fueled by the promise of developing polariton based spinoptronic devices, nonresonant excitation schemes seem the likely direction for future applications. In such a setup, optical or electrical excitation builds up a density of incoherent excitons in the cavity which eventually triggers bosonic stimulated scattering of polaritons with subsequent buildup of coherence and polarization as they condense~\cite{kasprzak_bose-einstein_2006}. This is in contrast to earlier works which relied on resonant excitation schemes to generate a coherent ensemble of polaritons~\cite{savvidis_angle-resonant_2000,lagoudakis_stimulated_2002, KavokinLagoudakisPRB2003}. There, interplay between linear and nonlinear phenomena (e.g., optical spin Hall effect and self-induced Larmor precession) has been studied both in the transient (pulsed excitation)~\cite{Renucci_PRB2005} and the continuous wave excitation regime~\cite{Tartakovskii_PRB2000, Gavrilov_PRB2016}. However, ideally, spinoptronic devices will either completely or partially operate using nonresonant excitation elements since it bypasses the need to fine tune the energy, momentum, and phase of a resonant laser. Moreover, such a device would likely operate well above condensation threshold in order to efficiently exploit the nonlinear spin dynamics of the polariton fluid. Nonlinearity is the needed ingredient for a device to perform nontrivial tasks~\cite{Amo_NatPho2010, Dreismann_NatMat2016}, but it can also destabilize the spin state of the condensate, affecting said device performance. Recent studies have mapped out interesting condensate regimes of polarization buildup, collapse, inversion, and hysteresis~\cite{Ohadi19}, as well as deterministic control of linearly polarized emission~\cite{Klaas_PRB2019}, but with most of the work focused on few selected polarization components. Therefore, a full characterization of the emission properties of optically confined polariton condensates, and its stability properties, is still lacking.

In this paper, we perform full-Stokes polarimetry of an optically trapped polariton condensate under nonresonant excitation~\cite{askitopoulos_polariton_2013, Cristofolini_PRL2013, askitopoulos_robust_2015}. We investigate the formation and depression of strongly polarized condensate regimes as a function of excitation beam polarization, power, and confinement area. We find that an interplay between the strain induced in-plane polarization splitting of the photonic mode, and condensate overlap with the non-condensed background of particles (referred as the {\it reservoir}) can destabilize the polariton pseudospin, resulting in a decrease of emitted polarized light. The polariton pseudospin describes the polarization of the condensate~\cite{Kavokin_PRL2004} just like the Stokes vector for an electromagnetic field and the two are explicitly related through the spontaneous emission of cavity polaritons whose polarization information is transferred with the escaping photons. Notably, under linearly polarized excitation we observe, with increasing excitation power, a transition from an unpinned- to a polarization pinned condensate~\cite{Read_PRB2009}. Our observation directly confirms the critical role of particle nonlinearity in pinning the pseudospin of the condensate. We additionally predict regimes of condensate limit cycles where an elliptically polarized pump gives rise to a stable self-induced Larmor precession of the condensate pseudospin.
\begin{figure}[t!]
    \centering
    \includegraphics[width=0.9\columnwidth]{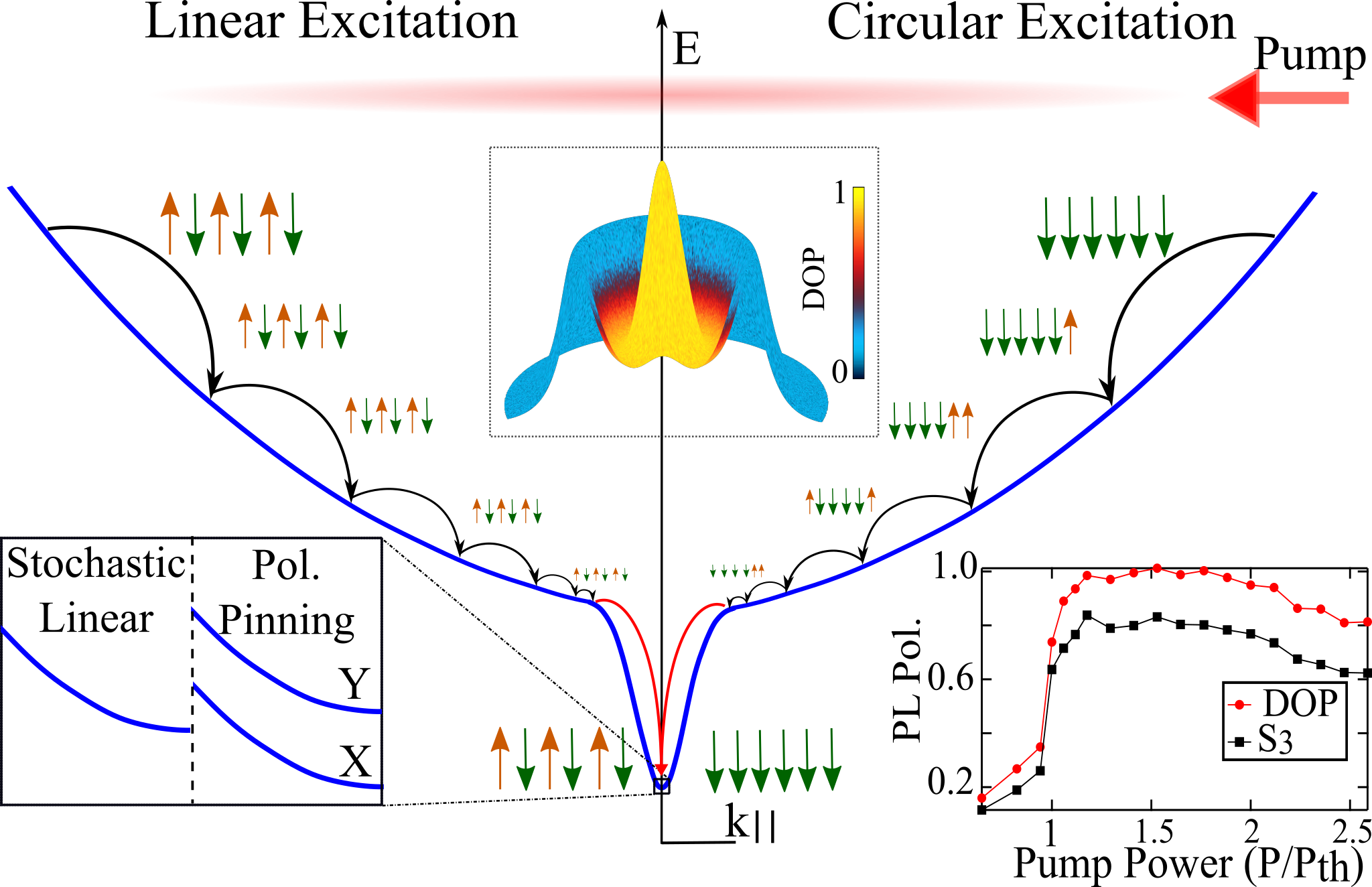}
    \caption{Schematic of the lower polariton dispersion (blue curve) illustrating the formation mechanism of the spinor condensate for circular (right side) and linear (left side) polarized nonresonant excitation (red area). Right inset: Measured power dependence of the optically trapped polariton condensate $S_3$ under circularly polarized pump and total degree of polarization (DOP). Left inset: Presence of birefringence splits the degeneracy between $X$ (horizontal) and $Y$ (vertical) polarized modes of the cavity field leading to pinning of the condensate pseudospin into the lower energy mode. Central inset, schematic representation of the emission at threshold colorcoded with DOP.}
    \label{fig1}
\end{figure}

\section{Experiment and results}
The sample used is a $2 \lambda$ GaAs microcavity with embedded InGaAs quantum wells, where we have previously demonstrated strong coupling and polariton condensation~\cite{cilibrizzi_polariton_2014}, held in a cryostat at $\unit[4]{K}$. The experiments are conducted in the strong coupling regime at negative exciton-cavity detuning ($\Delta\approx-3$ meV). The sample is excited nonresonantly ($\lambda=\unit[783.6]{nm}$) with a single mode continuous wave optical source, that is time-modulated at $\unit[1]{kHz}$ and 1\% on-off ratio with an accousto-optic modulator, and spatially modulated with a phase only spatial light modulator [see Secs.~S1-S3 in Supplementary Material (SM) for details on the experimental configuration].

Typical electron spin relaxation timescales in GaAs based systems are much longer than the relaxation time to the excitonic mode~\cite{Shelykh_semiconductorPRB2004}. As such, some of the degree of circular polarization of the continuous wave beam is transferred to the spin populations of the incoherent excitonic reservoir (see black arrows in Fig.~\ref{fig1}). Above the polariton condensation threshold, spin-conserving stimulated scattering of polaritons from the reservoir starts forming a mostly cocircularly polarized condensate (red lower arrows in Fig.~\ref{fig1}). In the case of linearly polarized excitation, and no pinning potential, the polarization of the condensate becomes randomly oriented on the $(S_1,S_2)$ equatorial plane of the Poincar\'e sphere since no specific condensate phase is adopted from the incoherent reservoir~\cite{Shelykh_PRL2006, baumberg_spontaneous_2008, ohadi_spontaneous_2012}. However, if the structure is anisotropic due to fabrication or strain, a finite linear polarization splitting can form which pins the condensate polarization (see left inset in Fig.~\ref{fig1})~\cite{Klopotowski_SSC2006, kasprzak_build_2007, balili_bose-einstein_2007, Read_PRB2009, Klaas_PRB2019}. The middle inset in Fig.~\ref{fig1} schematically shows the pumped reservoir (blue color) and the trapped condensate (yellow color) where the colorscale denotes the condensate's degree of polarization (the height is arbitrary).

The polarization of the emitted light provides direct access to the condensate spin, since the two circular polarizations couple independently with spin-up and spin-down optically active exciton states in the quantum wells through angular momentum selection rules. In their condensed form, polaritons can be expressed by a spinor order parameter $\Psi = (\psi_+, \psi_-)^T$ with spin-up and spin-down polaritons $(\psi_\pm)$ corresponding to right- and left-circularly polarized light respectively. Polariton spin physics are often conveniently described in the pseudospin formalism~\cite{Kavokin_PRL2004} as the polarization of the emitted light relates explicitly to the polariton spin structure such that the Stokes vector $\mathbf{S}$ is a measure of the polariton pseudospin. With the total particle number in the condensate written $S_0 = |\psi_+|^2 + |\psi_-|^2$, the normalized components of the Stokes vector $\mathbf{S} = (S_1,S_2,S_3)^T$ are written $S_1 = 2\text{Re}{(\psi_-^* \psi_+)}/S_0$, $S_2 = -2\text{Im}{(\psi_-^* \psi_+)}/S_0$, and $S_3 = (|\psi_+|^2 - |\psi_-|^2)/S_0$. 

\begin{figure}[t!]
    \centering
    \includegraphics[width=0.99\columnwidth]{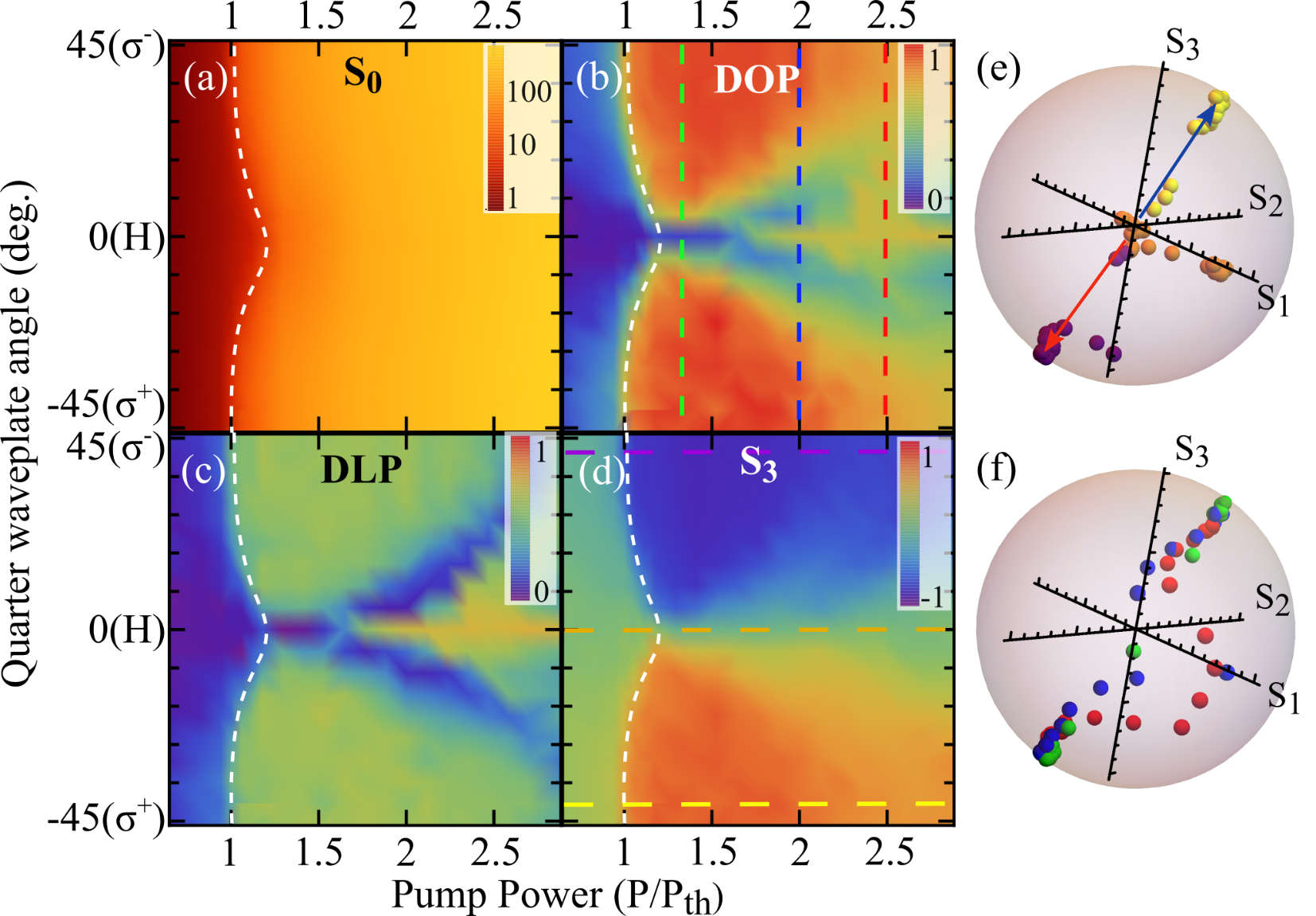}
    \caption{(a) Total emission intensity $S_0$ in arbitrary units, (b) DOP, (c) DLP, and (d) $S_3$ as a function of pump power $P$ and ellipticity (QWP angle) for a ring excitation geometry with a diameter of $d = 12$ $\mu$m. QWP angles $\mp45^\circ$ and $0^\circ$ correspond to right-, left-circular, and linear polarized excitation. White dashed lines mark the condensation threshold. (e,f) Three dimensional representation of the Stokes components with different colors corresponding to the same-color horizontal and vertical dashed lines in (d,b).}
\label{fig2}
\end{figure}

Our measurements are conducted entirely in the regime where condensation of polaritons occurs always in the ground state of the optically induced trap at low momenta and no higher order modes are excited for the full range of condensate densities studied~\cite{askitopoulos_robust_2015}. In order to filter out residual emission from the reservoir and collect only the condensate PL, we perform $k$-space filtering of wave vectors more than $\pm 1$ $\mu$m$^{-1}$. Figure~\ref{fig2} shows the time-integrated polarization from a trapped polariton condensate for a pump ring of diameter $d = 12$ $\mu$m as a function of both pump power and ellipticity controlled by rotation of a quarter wave plate (QWP) in the excitation path. Here, QWP $= \mp45^\circ$, and $0^\circ$ correspond to right-, left-circular, and linear polarization of the excitation laser respectively. Between those values the pump is elliptically polarized. Figures~\ref{fig2}(a-d) show the condensate total emission intensity $S_0$, degree of linear polarization $\text{DLP} = \sqrt{S_1^2 + S_2^2}$, total degree of polarization $\text{DOP} = \sqrt{S_1^2 + S_2^2 + S_3^2}$, and circular polarization $S_3$ respectively. Figure~\ref{fig2}(a), shows the threshold behavior of the condensate PL marked by the white dashed line, revealing that the condensation threshold is higher for linearly polarized excitation (around $1.18P_\text{th}$) as opposed to a circular polarized excitation which we define as $P_\text{th}$. This effect appears due to the same-spin Coulomb exchange interactions dominating over opposite spin interactions~\cite{Inoue_PRB2000, vladimirova_polariton-polariton_2010}. A right (left) circular polarized excitation beam results in a more spin up (down) populated reservoir of incoherent excitons which will sooner reach threshold density and undergo stimulated scattering into a cocircularly polarized condensate [see Fig.~\ref{fig2}(d)]~\cite{savvidis_angle-resonant_2000, lagoudakis_stimulated_2002, Gavrilov_APL2013}. In Fig.~\ref{fig2}(b) we observe that below threshold, marked by the white dashed line, the absence of stimulation mechanisms results in unpolarized PL emission with DOP close to zero. Above threshold, a sharp increase in the DOP marks the formation of the condensate order parameter with emission almost fully polarized. The notable exception is for linear polarization of the pump (QWP $= 0^\circ$), where the DOP still remains zero just above threshold (between $1.28 P_\text{th}$ and $1.59P_\text{th}$). In Sec.~S4 in the SM we provide details on the $S_{1,2}$ Stokes components.

For higher excitation powers and $\approx 99\%$ linear degree of polarization of the pump (QWP $\sim 0^\circ$), we observe the formation of a linear polarization `island' [see Fig.~\ref{fig2}(c)], which is attributed to an interplay of in-plane polarization splitting (due to sample strain/birefringence) and increased condensate nonlinearity~\cite{Read_PRB2009} leading to pinning of the condensate pseudospin. The absence of this linearly polarized island at lower powers is due to low occupation of the condensate (small nonlinearity) making it weakly pinned and, to our knowledge, has not been reported before. Stochastic noise then instead sets the pseudospin on a random walk on the Poincar\'e sphere (see Fig.~S7 in SM for details). Past experiments have shown either the immediate buildup of a pinned polarization above threshold~\cite{kasprzak_build_2007}, shot-to-shot stochastic polarization~\cite{baumberg_spontaneous_2008,ohadi_spontaneous_2015}, or $S_3$ spin flips~\cite{redondo_stochastic_2018}. Our observation implies that pinning cannot occur unless large enough particle numbers in the condensate are achieved (i.e., at higher powers). Figures ~\ref{fig2}(e,f) show the condensate Stokes components on the Poincar\'e sphere corresponding to the colored dashed lines in Fig. \ref{fig2}(d,b).

Formally, splitting between the polariton pseudospin components can be described by an effective magnetic field $\boldsymbol \Omega(\mathbf{r}) = (\Omega_x,\Omega_y,\Omega_z)$ which rotates the condensate pseudospin. Here, the $z$ direction is taken along the crystal growth axis, normal to the cavity plane. The corresponding Hamiltonian, in the basis of $\psi_\pm$, can be written $\mathcal{H}_\Omega(\mathbf{r}) = \hbar \boldsymbol \Omega(\mathbf{r}) \cdot \boldsymbol\sigma$ where $\boldsymbol \sigma$ is the Pauli matrix vector. Since the in-plane birefringence $(\Omega_x,\Omega_y)$ is random across the sample, the magnetic field is coordinate, $\mathbf{r}$, dependent. Correspondingly, we observe that the polarization of the pinning island strongly depends on the location of the excitation spot (see Sec.~S5 in SM).
 \begin{figure}[t!]
    \centering
    \includegraphics[width=0.8\columnwidth]{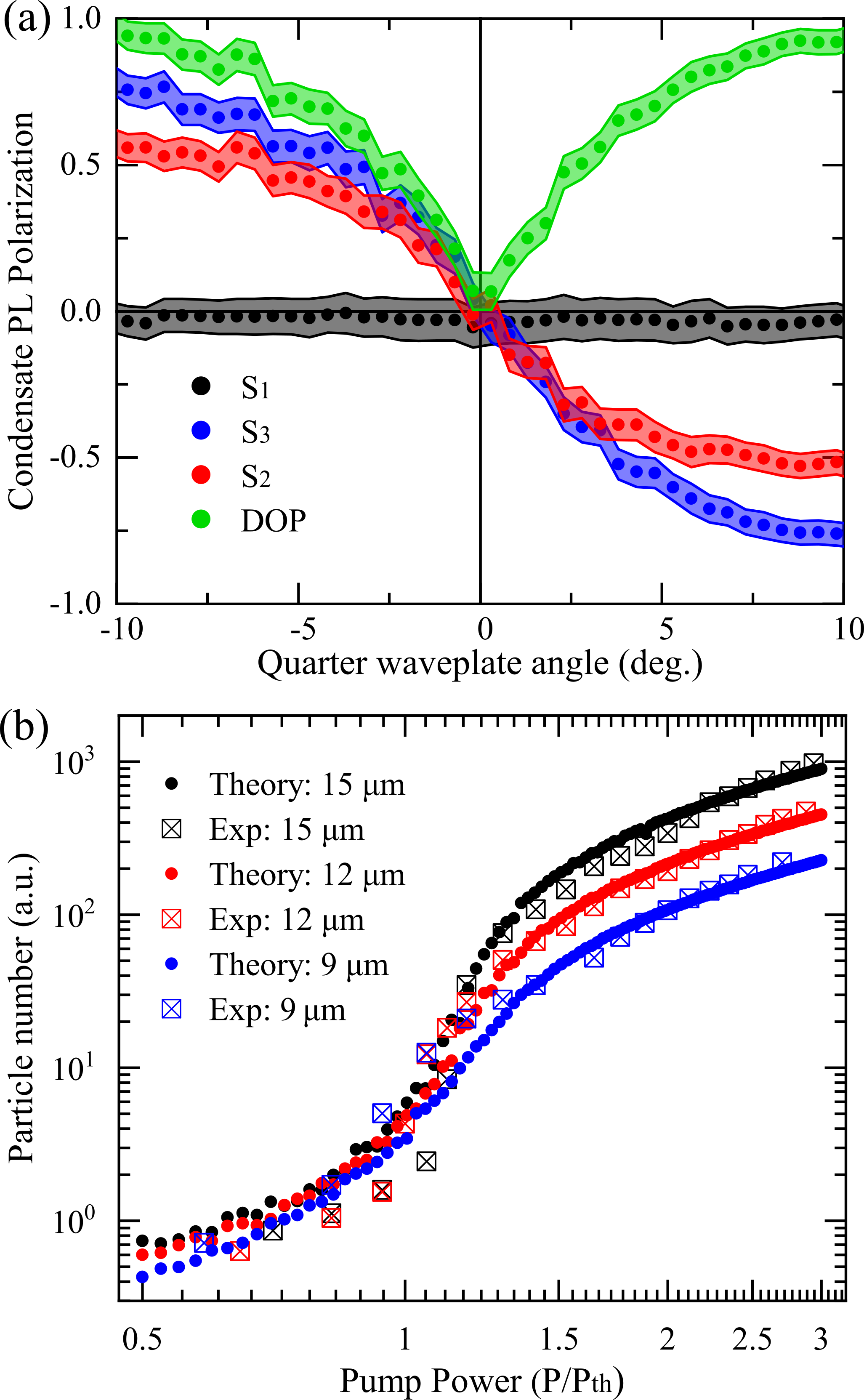}
    \caption{(a) Condensate polarization for a ring pump of diameter $d=12$ $\mu$m at $P = 1.23P_\text{th}$, fine-resolved for a pump laser polarization from QWP = -10$^\circ$ to 10$^\circ$. (b) Measured integrated photoluminescence (squares) and simulated $\bar{S}_0$ (dots) for different trap diameters and QWP $ = -45^\circ$. Horizontal axis is given in $P_\text{th}$ corresponding to a $d = 12$ $\mu$m pump.}
    \label{fig3_2}
\end{figure}

For the $S_3$ Stokes component, shown in Fig.~\ref{fig2}(d), the condensate circular polarization follows the handedness of the pump as expected~\cite{Shelykh_semiconductorPRB2004, ohadi_spontaneous_2012, Gavrilov_APL2013, Klaas_PRB2019}. Interestingly, even a very small ellipticity (QWP$ \approx \pm 2^\circ$) of the optical excitation is enough to set the condensate circular polarization [see Fig.~\ref{fig3_2}(a)]. The spin-imbalanced condensate and background reservoir of uncondensed polaritons, denoted $X_\pm$, result in an effective out-of-plane magnetic field, $\Omega_z = \alpha(|\psi_+|^2 - |\psi_-|^2) + g(X_+ - X_-)$ due to polariton-polariton interactions $\alpha$, and polariton-reservoir interactions $g$. This can cause the condensate pseudospin to start precessing around this interaction induced out-of-plane magnetic field which suppresses the $S_{1,2}$ components in the time-average measurements (i.e., regimes of almost zero DLP but finite $S_3$). For increasing beam ellipticity the condensate starts to become more pinned along the stronger $\Omega_z$ magnetic field, observed as an increase in $S_3$ [see Fig.~\ref{fig2}(d) and Fig.~\ref{fig3_2}(a)]. We note that similar results were obtained  for detunings between $-4$ meV to $-2$ meV (see Sec.~S6 in SM).
 \begin{figure}[t!]
	\centering
	\includegraphics[width=1\columnwidth]{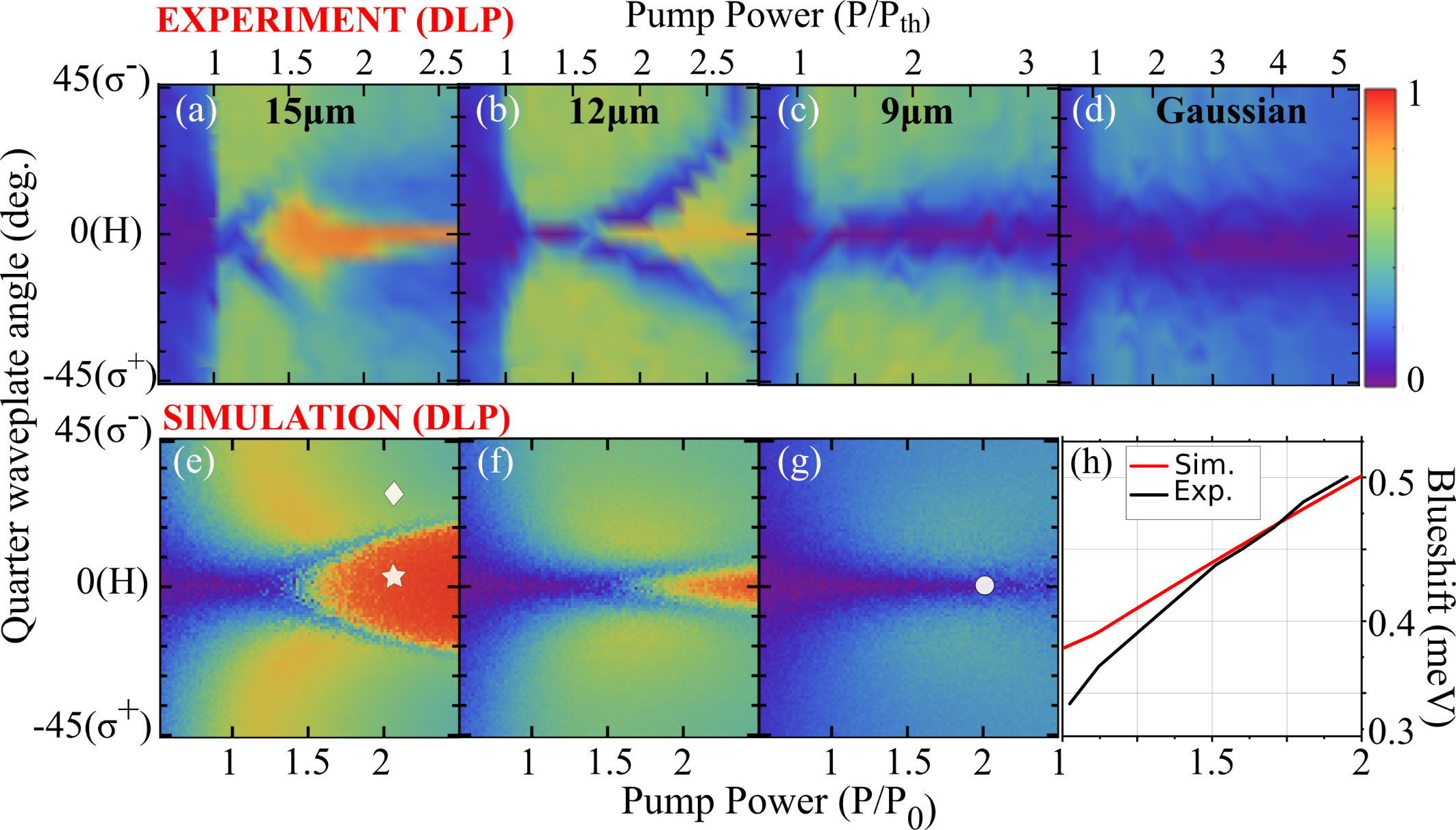}
	\caption{DLP polarization maps as a function of pump power and QWP angle. Panels (a-c) show results for decreasing pump diameter $d = 15, 12, 9$ $\mu$m. (d) Shows the condensate DLP in the case of a Gaussian spot excitation (no confinement). (e-f) Simulated time-average condensate DLP using Eqs.~\eqref{eq.1}-\eqref{eq.3}. The star, circle and diamond markers correspond to panels (a-c) in Fig.~\ref{fig5}. (h) Black line shows the measured condensate blueshift at QWP $=0^\circ$ in a $d = 12$ $\mu$m pump trap for increasing power. Red line is the simulated energy of the condensate.}
	\label{fig3}
\end{figure}

Interestingly, the linearly polarized island is enclosed by two depolarized streaks [see Fig.~\ref{fig2}(b,c)] which, to the best of our knowledge have not been previously reported. These streaks correspond to the interface between the pseudospin being pinned either by the in-plane magnetic field $(\Omega_x,\Omega_y)$ from birefringence or the interactions-induced magnetic field $\Omega_z$. In-between these two pinning regimes the pseudospin is very sensitive to background white noise which can stochastically move it from precessing around one field to the other, causing the measured polarization to appear completely depolarized.

In Fig.~\ref{fig3_2}(b) we show the measured integrated condensate PL (squares) as a function of pump power, with QWP $=-45^\circ$, and for three different condensate trap sizes. We point out that the trap size corresponds to the diameter of the ring shaped excitation beam. Interestingly, we observe for smaller traps a decrease in the polariton condensate occupation number which we address in the next section. 

In Fig.~\ref{fig3} we show that the size of the condensate trap has a pronounced effect on the measured polarization patterns. By decreasing the diameter of the excitation ring, we observe that the aforementioned linear polarization island decreases [see Fig.~\ref{fig3}(b)] until it vanishes completely [see Fig.~\ref{fig3}(c)]. When the condensate is excited with a Gaussian excitation spot, such that it does not experience any optical confinement, the DLP decreases even more [see Fig.~\ref{fig3}(d)]. This observation demonstrates that increasing the overlap between the condensate and its background reservoir (i.e., smaller diameters) results in strong depolarization, as supported by numerical simulations detailed below. It should be noted that for the measurements with a Gaussian excitation spot most of the emitted light coincides with the spot area. Any light away from the spot, possibly adopting nontrivial polarization textures~\cite{kammann_nonlinear_2012, cilibrizzi_polariton_2015, cilibrizzi_PRB2016}, has a negligible contribution to our averaged measurements.
\begin{figure}[t!]
    \centering
    \includegraphics[width=1\columnwidth]{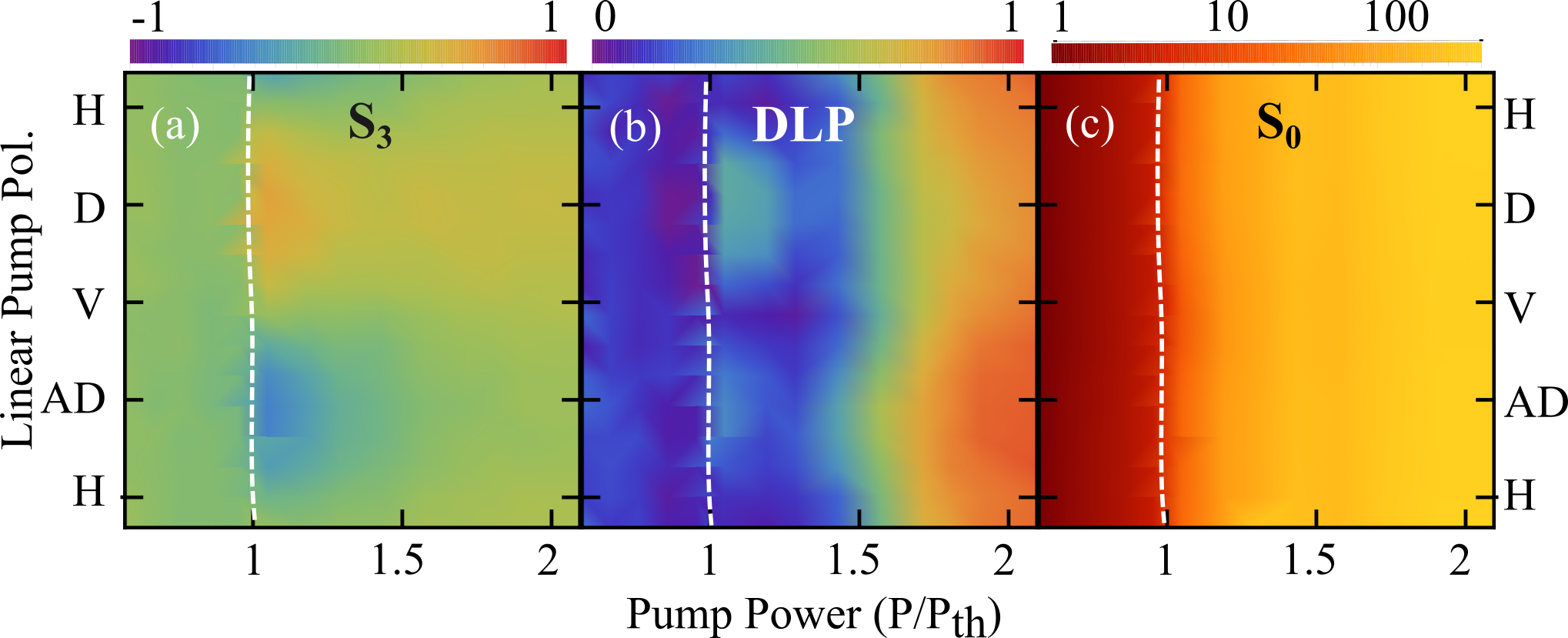}
    \caption{Condensate polarization as a function of pump power and incident linear polarization angle. (a) $S_3$, (b) DLP, (c) and total emitted intensity. The vertical axis denotes the half-waveplate angle and is marked with different linear polarizations. Horizontal (H), vertical (V), diagonal (D), and anti-diagonal (AD).}
    \label{fig4}
\end{figure}

By recording the condensate polarization under different linear polarization angles of the excitation, we find that the condensate $S_3$ and DLP at high powers is mostly invariant as a function of the linear polarization angle [see Fig.~\ref{fig4}(a,b)]. This further confirms that the pinning is not a result of transferred linear polarization from the excitation to the condensate. Moreover, the condensation threshold is uniform for all linear polarization angles of the excitation [see Fig.~\ref{fig4}(c)]. We point out that the condensate polarization varies slightly when excitation switches from diagonal to anti-diagonal polarization. Just above threshold, we observe a small formation of $S_3$ component with different signs that is suppressed as the condensate density increases [see Fig.~\ref{fig4}(a)]. We attribute this to a small pump ellipticity induced by the optical elements of our excitation setup as well as a non-zero retardation of the excitation beam due to sample birefringence (see Sec.~S7 in SM).

\section{Theory}
The destabilization of the condensate pseudospin and consequent depolarization of the system can be modeled through a set of driven-dissipative stochastic (Langevin-type) Gross-Pitaevskii equations~\cite{Carusotto_RMP2013} coupled to spin-polarized rate equations describing excitonic reservoirs $X_\pm = X^{A}_\pm + X^{I}_\pm$ feeding the condensate,
\begin{align} \notag
i \dot{\psi}_\sigma  = & \theta_\sigma(t) + \frac{1}{2} \Big[\alpha |\psi_\sigma|^2  + g (1 - \eta P_\sigma)   (X^{A}_\pm + X^{I}_\pm)  \\ \label{eq.1}
& + i \left( R (1 - \eta P_\sigma) X^A_\sigma - \Gamma \right) \Big] \psi_\sigma - \frac{\Omega_x}{2} \psi_{-\sigma}, \\ \notag
\dot{X}^A_\sigma  =&   - \left( \Gamma_A + R (1 - \eta P_\sigma) |\psi_\sigma|^2 \right)X^A_\sigma  \\
& + \Gamma_s(X^{A}_{-\sigma} - X^A_\sigma) + W X^I_\sigma  ,
 \\  \label{eq.3}
 \dot{X}^I_\sigma  = &  - \left( \Gamma_I + W\right)X^I_\sigma + \Gamma_s(X^{I}_{-\sigma} - X^I_\sigma) + P_\sigma.
 \end{align}
Here, we take into account the presence of spin-polarized active and inactive reservoirs $X^{A,I}_\sigma$ respectively~\cite{Lagoudakis2010a, Lagoudakis_PRL2011, Klaas_PRB2019}. The former satisfies energy conservation rules of particles scattering into the condensate whereas the latter does not. Here, $R$ is the spin-conserving rate of stimulated scattering of polaritons into the condensate, $\Gamma$ is the polariton condensate decay rate, $\Omega_x$ represents a birefringence induced effective magnetic field which splits the polariton $XY$ polarizations (see left inset in Fig.~\ref{fig1}), $\Gamma_{A,I}$ are the decay rates of active and inactive reservoir excitons respectively, $W$ is the conversion rate between inactive and active reservoir excitons, $\Gamma_s$ is a spin-flip rate of excitons in each reservoir, and $P_\pm = P_0 \cos^2{(\text{QWP} \pm \pi/4)}$ is the power of the nonresonant continuous wave pump. The parameter $\eta$ phenomenologically captures the sublinear dependence of the ground state energy shift and gain with increasing pump power. Such sublinear power dependence can be physically understood from the decreasing overlap between the more tightly confined condensate and the surrounding reservoir.

The correlators of the background shot noise from the reservoir $\theta_\pm(t)$ are written,
\begin{align}
\langle d\theta_\sigma(t) d\theta_{\sigma'}(t') \rangle & =  \frac{\Gamma + R X^A_\sigma}{2}  \delta_{\sigma \sigma'} \delta(t-t'),\\
\langle d\theta_\sigma(t) d\theta_{\sigma'}^*(t') \rangle & = 0.
\end{align}
We define a time-averaged Stokes component from simulation as $\bar{S}_3 = [\int_0^T |\psi_+|^2-|\psi_-|^2 dt] / \int_0^T S_0 dt$ where $T$ is the simulated time interval. The $\bar{S}_{1,2}$ components are calculated analogously. Below condensation threshold there is no buildup of a coherent polariton state and the simulated average pseudospin is zero, $\bar{\textbf{S}} = \mathbf{0}$. The threshold is defined as the point where gain and losses balance against each other, written $R X^A_{\pm} - \Gamma = 0$. Above threshold, gain overcomes losses and a coherent polariton state $\Psi$ forms.

Decreasing the pump diameter $d$ affects some parameters of Eqs.~\eqref{eq.1}-\eqref{eq.3}. Namely, $\alpha, \Gamma, g, R$. The polariton-polariton interaction strength $\alpha$ increases because of a decreased localization of the condensates which scales as $\alpha \propto \int |\Psi(\mathbf{r})|^4 d\mathbf{r}$~\cite{pethick_bose-einstein_2001}. Assuming that the condensate occupies the ground state of a cylindrically symmetric two-dimensional harmonic potential $V(r) = V_0 (2r/d)^2$, whose oscillator strength changes with pump diameter, we have $\Psi(\mathbf{r}) = \beta/\sqrt{\pi} \exp{[-\beta^2 r^2 / 2]}$ where $\beta = [8m V_0/\hbar^2 d^2 ]^{1/4}$ (where $m$ is the polariton mass). One then obtains that $\alpha \propto 1/d$. The change in overlap between the reservoir and the condensate $g,R \propto \int  P(\mathbf{r}) |\Psi(\mathbf{r})|^2 d\mathbf{r}$ is more challenging to estimate as it depends on the details of the pump shape $P(\mathbf{r})$. We find a good fit to the experimental results with the dependence $g,R \propto 1/d^3$. From Fig.~\ref{fig3}(a-c) it can be seen that the threshold of the condensate does not depend too strongly on the trap size. This can be attributed to an increase in the condensate decay $\Gamma$ due to the enhanced escape rate of the more energetic polaritons in smaller traps~\cite{Cristofolini_PRL2013}. We therefore adopt $\Gamma \propto 1 /d^3$ dependence such that the threshold condition, $R X_\pm^A - \Gamma = 0$, becomes invariant of trap diameter $d$. Parameters used in simulations are given in~\cite{parameters}.

Results from simulation are shown in Fig.~\ref{fig3}(e-g) where the $\text{DLP} = [\bar{S}_1^2 + \bar{S}_2^2]^{1/2}$ is plotted in comparison to experimental observations for varying pump diameters $d$, showing good agreement. Pump power in simulation is given in units of $P_0 =2 \Gamma\Gamma_A(\Gamma_I+W) / (RW)$ which is the threshold power for $P_+=P_-$. In Fig.~\ref{fig3}(h) we show the measured and simulated condensate blueshift as a function of pump power at QWP $=0^\circ$ and $d = \unit[12]{\mu m}$. We observe from simulation [see Fig.~\ref{fig3}(e)] that the DLP of the polarization island, after its formation, does not decrease with increasing pumping powers. Therefore, the stability of the island is not sensitive to the monotonically increasing blueshift experienced by the condensate when pump power is increasing. Instead, for QWP $ = 0^\circ$, the vanishing of the polarization island depends on the strength of the in-plane field $\Omega_x$ against the polariton-polariton nonlinearity. Indeed, as the trap diameter decreases the scattering rate $R\propto 1/d^3$ increases proportionally, leading to saturation of the reservoirs $X^A_\sigma$ for smaller particle number $\bar{S}_0$. Consequently, we observe in both experiment and simulation a strong decrease in $\bar{S}_0$ for smaller trap sizes [see Fig.~\ref{fig3_2}(b)].

This evidences that the nonlinearity of the condensate, which scales with particle number $\bar{S}_0$, is the crucial mechanism along with $\Omega_x$ in order to pin the polarization~\cite{Read_PRB2009}. This is in agreement with our experimental observations that the linearly polarized island could only form at higher pumping powers above threshold [see Fig.~\ref{fig2}(b,c)]. For elliptically polarized pumps (QWP $\neq 0^\circ$) the spin-imbalanced reservoir and condensate start playing a role in the vanishing of the island through the pump induced component $\Omega_z$ which tilts the net effective magnetic field out of the cavity plane. A simplified Gross-Pitaevskii model, detailed in the SM, verifies this interpretation of the island's destabilization. We note that the results reported here do not depend strongly on inclusions of spin-anisotropic interactions, nor the precise value of $\Gamma_s$. Details on individual simulated Stokes components are given in Secs.~S8 in SM.
\begin{figure}[t!]
	\centering
	\includegraphics[width=1\columnwidth]{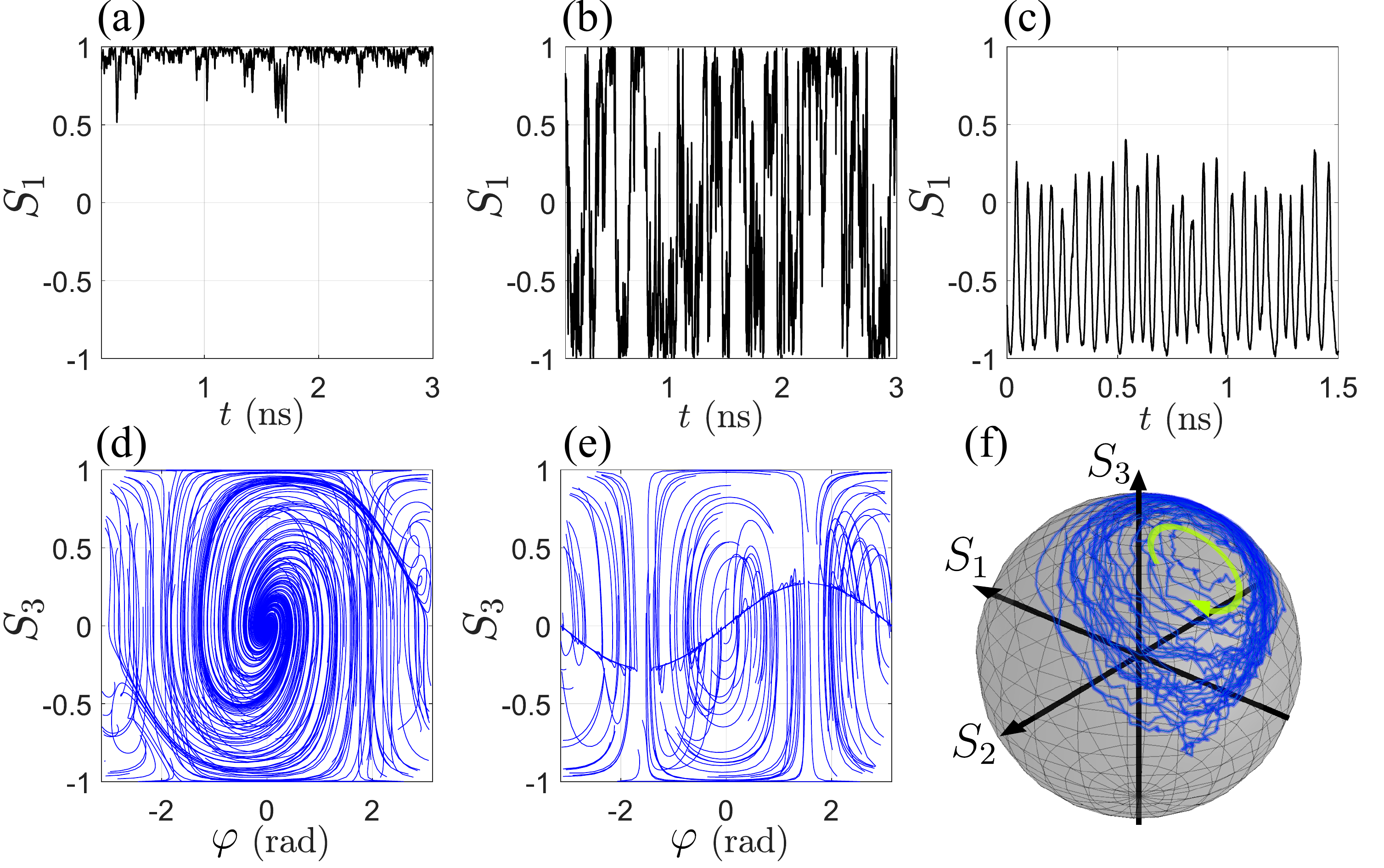}
	\vspace{-10pt}
	\caption{(a-c) Dynamics of the normalized $S_1$ component from the star, circle, and diamond markers in Fig.~\ref{fig3} respectively. (a) For large traps the pseudospin is pinned along the effective magnetic field $\Omega_x$. (b) For smaller pump diameters the condensate blueshifts and the pseudospin starts destabilizing and fluctuates between $S_1 = \pm1$. (c) For elliptical excitation one gets $|\Omega_z| \gtrsim |\Omega_x|$ which can set the condensate into a tilted limit cycle appearing as persistent oscillations in the Stokes components. (d,e) Overlaid pseudospin phase space trajectories with $\theta_\pm(t) =0$ but random initial conditions, corresponding to the star and circle in Fig.~\ref{fig3} respectively. $\varphi$ corresponds to the azimuthal angle of the Poincar\'e sphere. Results show a change from one dominant attractor to two weaker ones. (f) Representation of the limit cycle (precession) in (c) on the surface of the Poincar\'e sphere.}
	\label{fig5}
	\vspace{-15pt}
\end{figure}

In Fig.~\ref{fig5} we show simulated pseudospin dynamics for three different values of pump power and QWP angles. Figures~\ref{fig5}(a-c) show the $S_1$ component corresponding to the star, circle, and diamond markers of Fig.~\ref{fig3}(e,g) respectively. When the pump diameter is large, the overlap with the reservoir is small and we see strong pinning of the $S_1$ component along the direction of the in-plane magnetic field [Fig.~\ref{fig5}(a)]. For smaller pump diameter [Fig.~\ref{fig5}(b)], the overlap between the condensate and reservoir increases causing the pseudospin to destabilize and start to stochastically fluctuate between $S_1 = \pm1$. In Fig.~\ref{fig5}(d,e) we investigate this process of destabilization by plotting overlaid phase space trajectories for large and small pump diameters ($d = 15$, and $9$ $\mu$m respectively). Here we set $\theta_\sigma(t) = 0$ but use random initial conditions for the integration of Eqs.~\eqref{eq.1}-\eqref{eq.3}. For large pump diameters there exists a dominant phase space attractor at $S_1 = 1$ ($\varphi = 0, \ S_3 = 0$), whereas for smaller diameters this attractor decreases and a second attractor forms around $S_1 = -1$ ($\varphi = \pi, \ S_3 = 0$). As such, stochastic fluctuations in the dynamics of the pseudospin start shifting the polarization randomly between $S_1 = \pm1$ causing the polarization island to vanish in the time-averaged measurements. For an elliptically polarized pump, the pseudospin undergoes a mixture of pump- and self-induced Larmor precession [see Fig.~\ref{fig5}(c,f)] overcoming the pinning potential $|\Omega_z|\gtrsim|\Omega_x|$, setting the condensate into a tilted limit cycle which manifests in our measurements as an effective depolarization. As predicted by previous theoretical studies~\cite{Read_PRB2009} the precession rotational axis becomes tilted towards the negative $S_1$ axis on the Poincar\'e sphere (opposite the pinning field).

\section{Conclusions}
We have experimentally investigated and analyzed the polarization characteristics of optically trapped polariton condensates. For the case of a circularly polarized excitation we have shown a sharp (up to DOP $ \approx 1$) increase of the degree of polarization above condensation threshold. The high polarization is consistent with recent observations on the extreme long coherence times of optically trapped polariton condensates~\cite{Askitopoulos_coherence_2019arXiv}. The condensate spin properties in these conditions are governed by optical orientation of the reservoir excitons mostly adopting the polarization of the optical excitation.

For a linearly polarized excitation, depending on the size of the pump induced confining potential and pumping power, we observe a transition from a regime of pinned linear polarization of the condensate to a completely depolarized condensate. The effect is attributed to competition between a sample dependent in-plane polarization splitting and overlap between the condensate with its reservoir. The former, in conjunction with increasing condensate nonlinearity, causes the appearance of a highly linearly polarized condensate due to a pinning effect. When the pump diameter decreases overlap with the reservoir increases and we observe rapid depolarization (unpinning) of the condensate due to weakening of its phase space attractor with increased stochastic spin fluctuations. We also report on the presence of limit cycles in the condensate pseudospin which contribute to the observed depolarization of the emission. Our results pave the way towards generating highly polarized polariton condensates using only nonresonant excitation techniques, promising for spin-dependent optoelectronic devices.

Interestingly, a previous report has shown that a random massive degree of circular polarization can build up for QWP $=0^\circ$ under very similar conditions as presented here (i.e., optical trapping of polaritons)~\cite{ohadi_spontaneous_2015,redondo_stochastic_2018}. There, spin-flip scattering of polaritons into the condensate from the reservoir was treated equal to spin-conserving scattering. In our work, the spin-flip rate $\Gamma_s$ between the reservoir components in our model contributes to such mixed scattering of polaritons but we do not observe such circularly polarized states. This begs the question whether there exist conditions where the regimes of pinned linearly polarized condensates and bifurcated circularly polarized condensates can be brought together to exploit nonresonant multistable operation on the Poincar\'e sphere in contrast to resonant schemes~\cite{Gippius_PRL2007, Paraiso_Nature2010,cerna_ultrafast_2013}.

\section*{Acknowledgements}
The authors acknowledge the support of the Skoltech NGP Program (Skoltech-MIT joint project), the UK Engineering and Physical Sciences Research Council (grant EP/M025330/1 on Hybrid Polaritonics) and by RFBR according to the research projects No. 20-52-12026 (jointly with DFG) and No. 20-02-00919.

\setcounter{equation}{0}
\setcounter{figure}{0}
\setcounter{section}{0}
\renewcommand{\theequation}{S\arabic{equation}}
\renewcommand{\thefigure}{S\arabic{figure}}
\renewcommand{\thesection}{S\arabic{section}}
\onecolumngrid

\vspace{1cm}
\begin{center}
\Large \textbf{Supplemental Material}
\end{center}

\section{Experimental setup}
\begin{figure}
    \centering
    \includegraphics[width=0.7\columnwidth]{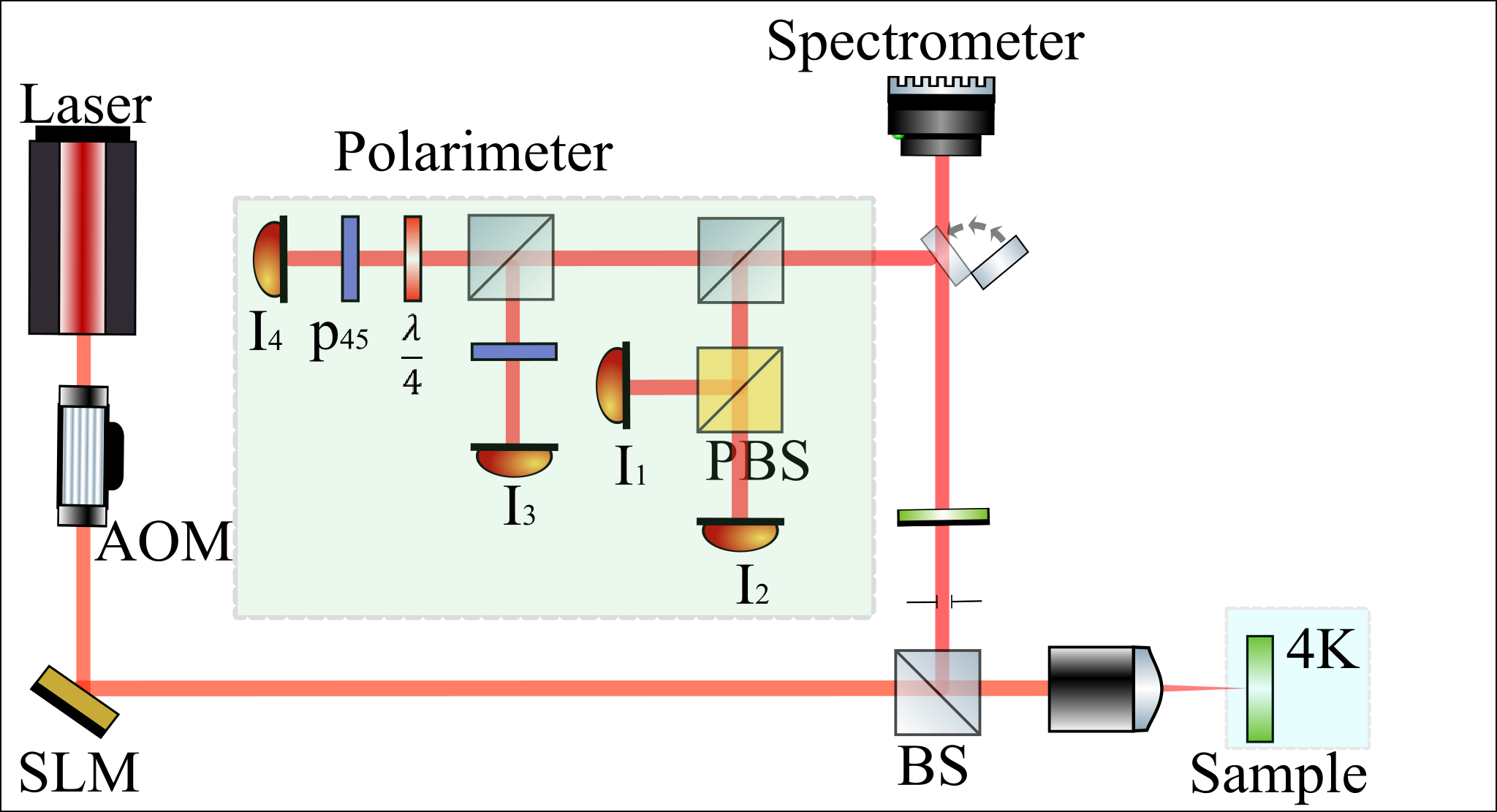}
    \caption{Schematic of the experimental setup and polarimeter. AOM, SLM, PBS, BS stand for acousto-optic modulator, spatial light modulator, polarized and non-polarized beamsplitter respectively.}
    \label{setup}
\end{figure}

Figure~\ref{setup} depicts a simplified schematic of the experimental setup. An $M^2$ SolsTiS  CW mono-mode laser is used to excite the sample by pumping at the first Bragg minimum at \unit[783.6]{nm}. The experiments are performed in a quasi-CW excitation regime. Namely, the pump laser radiation was modulated with an acousto-optic modulator with 1 kHz frequency and 1$\%$ duty-cycle (10 $\mu$s pulse length) to diminish heating of the sample. Then, a phase only spatial light modulator transforms the Gaussian beam into a ring-shaped profile, which creates the optical confining potential for polaritons. We investigate the photoluminescence (PL) of the sample in a confocal configuration using a 50x objective with 0.42 NA for excitation and collection of the sample PL.

We detect real- and Fourier-space images of the sample PL. Also, we characterize the spectrum and dispersion of the PL by using a Princeton Instruments \unit[750]{mm} Spectrometer with a Pixis CCD. For the polarization characterization of the polariton condensate, we use a specially developed full-Stokes polarimeter (see Sec.~\ref{sec.Stokes}). Since PL intensity is quite low, especially below the condensation threshold, we use lock-in amplifiers together with Si-photodiodes to measure polarization power dependence. The lock-in integration time is set to 100 ms that results in integration of the signal over approximately 100 pulses.

\section{Stokes parameters measurement} \label{sec.Stokes}
The polarimeter (see Fig.~\ref{setup}) is based on the division of amplitude approach. Firstly, a beamsplitter (BS) 50:50 divides incoming light into two equal parts. One of them falls onto a polarizing beamsplitter, and the intensities of horizontal $I_1$ and vertical $I_2$ components are registered by two photodetectors.

The remaining half of the light hits a second BS and splits into two equal parts. One part goes to a third detector ($I_3$) through the polarizer (p$_{45}$) having transmission axis rotated at 45 degrees to detect the diagonally polarized intensity. The other quarter of light is used to detect the right circular polarization through an inversed circular polarizer, namely a quarter waveplate $\lambda/4$ and a polarizer oriented at 45 degrees.

Having obtained readings from the four detectors we have:
\begin{equation}
I_H = 2I_1, \qquad I_V = 2I_2, \qquad  I_D = 4 I_3, \qquad I_{\sigma^+} = 4 I_4,
\end{equation}
where $I_{H,V,D,A,\sigma^+,\sigma^-}$ denote the intensity of light belonging to horizontal, vertical, diagonal, anti-diagonal, right circular, and left circular polarized light, respectively. We can then reconstruct the normalized Stokes vector as follows, 
\begin{gather}\label{StokesVectorPolarimeter}
    S_1=  \frac{I_H - I_V}{I_H + I_V}= \frac{I_1-I_2}{I_1+I_2},\nonumber\\
    S_2=\frac{I_D - I_A}{I_D + I_A}=\frac{4I_3-(I_1+I_2)}{I_1+I_2},\\
    S_3= \frac{I_{\sigma^+} - I_{\sigma^-}}{I_{\sigma^+} + I_{\sigma^-}}=\frac{4I_4-(I_1+I_2)}{I_1+I_2}.\nonumber
\end{gather}
The polarimeter is carefully calibrated with laser light of a known polarization tuned to the condensate emission wavelength in order to account for any imperfections (i.e. the beam splitter is not exactly 50/50 for all polarizations) of the optical elements used. The developed optical scheme can detect fast polarization changes. The speed of data collection is limited just by the rise- and dead- times of the photodetectors. However, in our experiment, the limiting factor of the polarimeter operation speed is the integration time (\unit[100]{ms}) of the lock-in amplifiers.

\section{Dispersion and condensation}
\begin{figure}
    \centering
    \includegraphics[width=0.9\columnwidth]{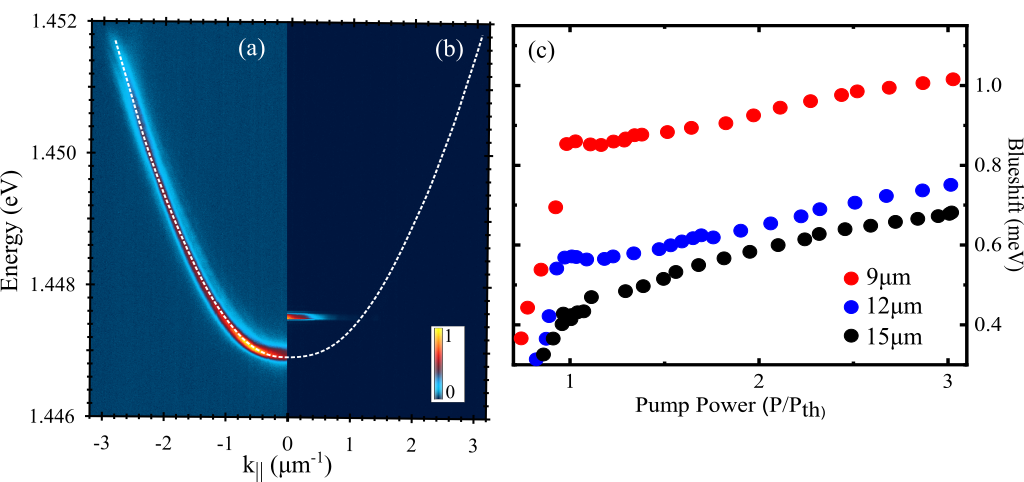}
    \caption{Energy resolved cavity PL at excitation powers (a) below and (b) 20\% above the condensation threshold. (c) Blueshift of the condensate energy as a function of pump power for different trap sizes relative to the free polariton energy.}
    \label{dispersions}
\end{figure}

We excite the sample nonresonantly at \unit[783.6]{nm}. Below the condensation threshold, we see the lower polariton branch on our imaging spectrometer. The bottom of the branch is at \unit[857]{nm} [see Fig.~\ref{dispersions}(a)]. Further increasing the intensity of the nonresonant excitation laser we observe blueshift of the lower polariton branch which is accompanied by condensation at in-plane wave vectors close to zero [see Fig.~\ref{dispersions}(b)]. In Fig.~\ref{dispersions}(c) we show the blueshift of the polariton condensate as a function of excitation power for different trap sizes. Above threshold the blueshift is increasing approximately linearly with power for all sizes and only differs by a constant energy shift. This is mostly due to the ground state trap level shifting to higher energy as the particle confinement ``tightens''. 

\section{DOP and Stokes components of condensate PL for different excitation ring sizes}
\begin{figure}
    \centering
    \includegraphics[width=0.7
    \columnwidth]{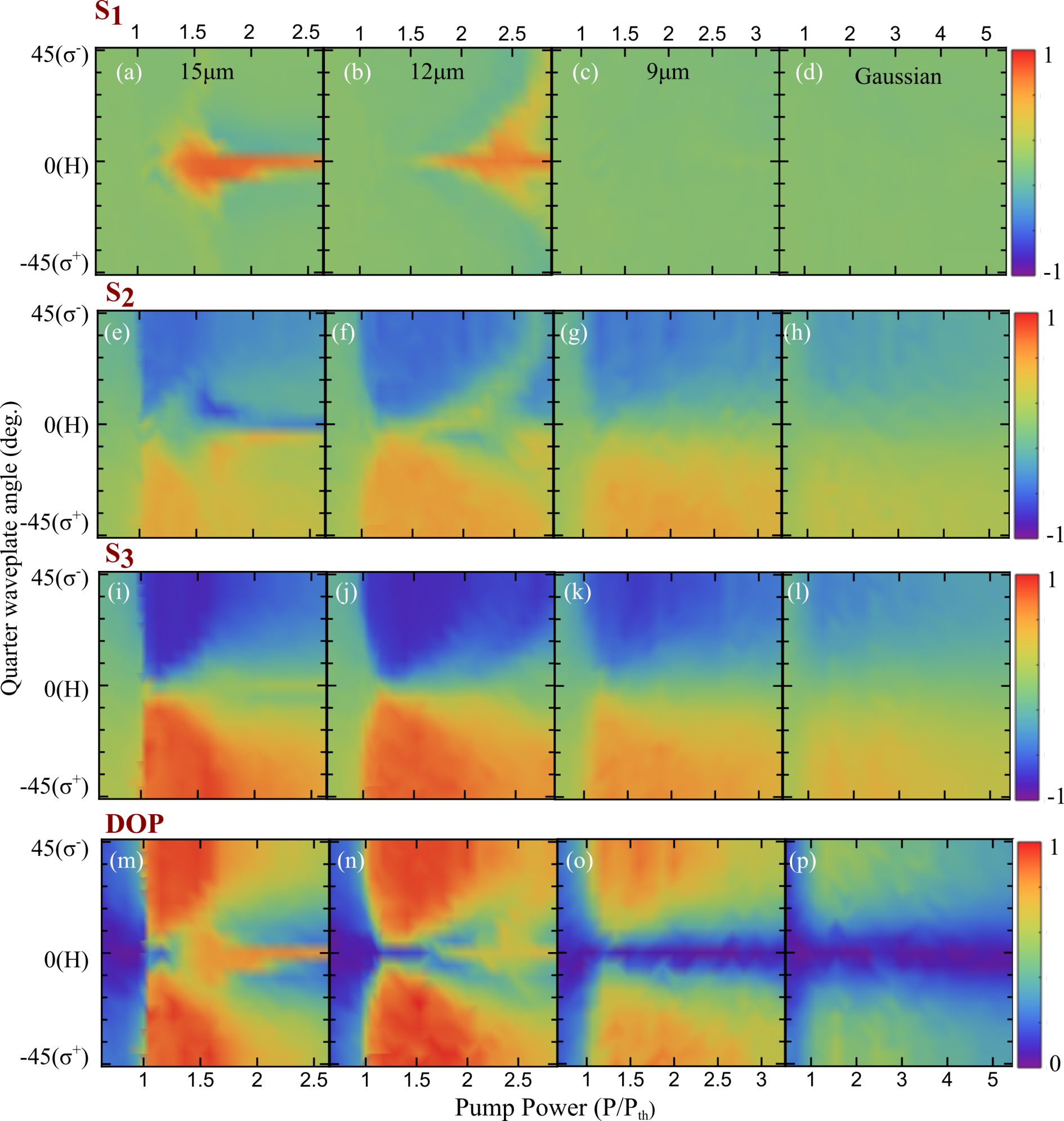}
    \caption{$S_1$ (a-c),  $S_2$ (e-h), $S_3$ (i-l), and DOP (m-p) for a ring excitation geometry of diameter $d = 15,12,9$ $\mu$m and a Gaussian excitation geometry of FWHM = 4 $\mu$m respectively.}
    \label{sizeDoP}
\end{figure}

Figure~\ref{sizeDoP} is the same experimental data as shown in Fig.~4 in the main text. It shows all Stokes components and the total degree of polarization (DOP) for the three studied excitation ring diameters $d = 15,12,9$ $\mu$m and a Gaussian pump of full-width-half-maximum (FWHM) = 4 $\mu$m.

For all sizes of the pumping ring and Gaussian excitation, we observe that the circular polarization of the condensate follows the circular polarization of the pump [see Fig.~\ref{sizeDoP}(i-l)]. Also, for all pumping configurations we observe gradual depolarization of the PL with increasing pump power. Under approximately linearly polarized excitation (marked with ``H'' on the vertical axis) we observe a transition from unpolarized PL to strongly linearly polarized pinning regime above the condensation threshold for larger ring diameters [see Fig.~\ref{sizeDoP}(a,b)]. Moreover, in this pinning regime the polarization of the linear polarization island is mostly independent of the excitation geometry [see Fig.~\ref{sizeDoP}(a,b) where the orange colored region appears for both cases].

\section{Excitation place dependence}
\begin{figure}
    \centering
    \includegraphics[width=0.45\columnwidth]{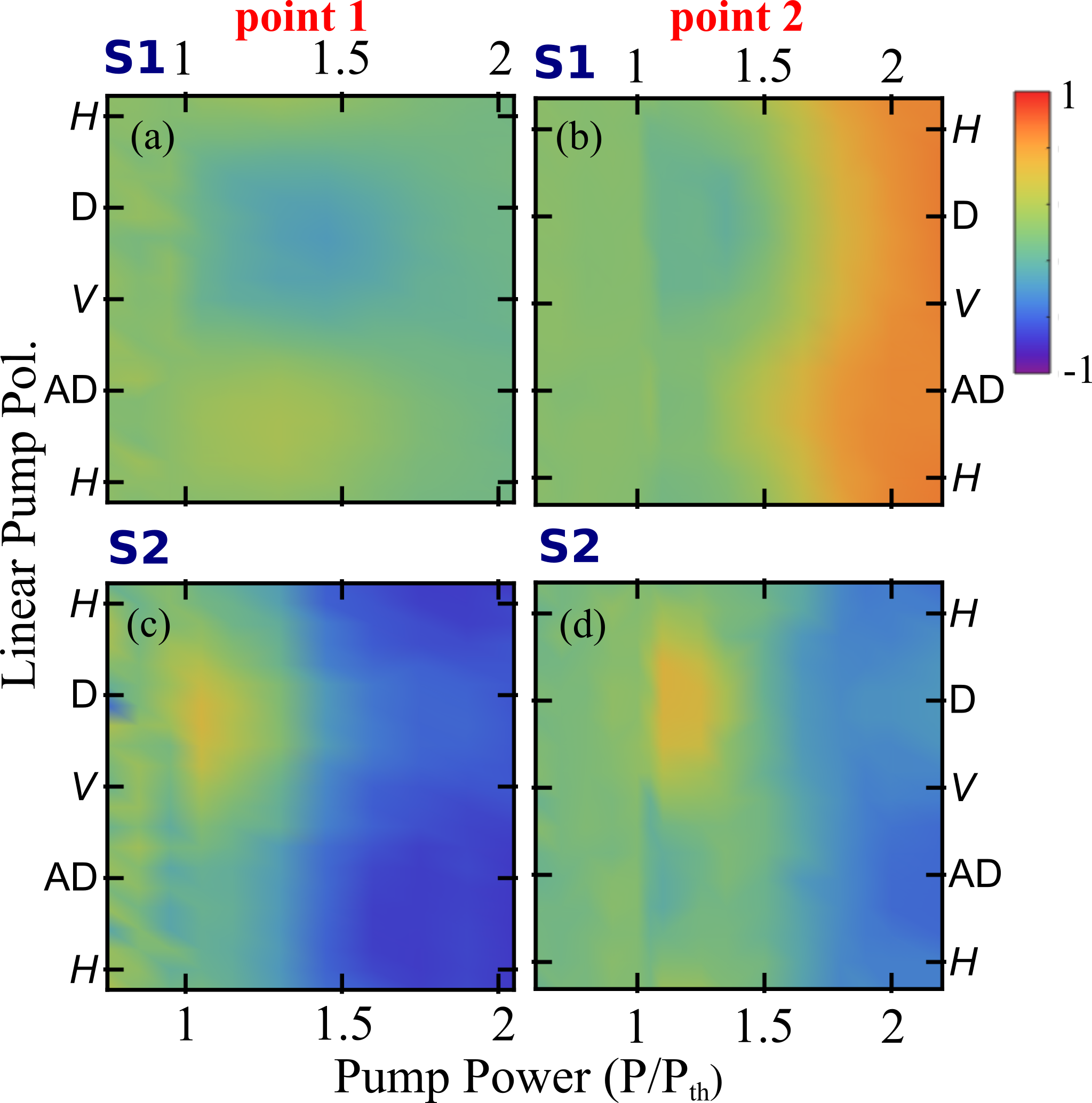}
    \caption{(a,b) $S_1$ and (c,d) $S_2$ polarization maps for two different sample positions for varying linear polarization of the excitation and power. }
    \label{point}
\end{figure}

Here we present polarization maps for $S_1$ and $S_2$ Stokes components of the polariton PL for two different points on the sample. For the first point [see Fig.~\ref{point}(a,c)] at high pump powers ($P>1.4 P_\text{th}$) we see an increase in the $S_2$ component, while the $S_1$ stays close to zero. On the other hand, for the second point [see Fig.~\ref{point}(b,d)] both $S_1$ and $S_2$ component obtain high values for large pump power. The results underline the position dependence of the sample birefringence which dictates the polarization of the pinned condensate. In other words, the effective in-plane polariton magnetic field $[\Omega_x(\mathbf{r}), \Omega_y(\mathbf{r})]$ is a position dependent random valued vector field. Moving from point to point on the sample, we can observe any mixture of linear and diagonal polarization pinning.

\section{Detuning dependence}
\begin{figure}
    \centering
    \includegraphics[width=0.45\columnwidth]{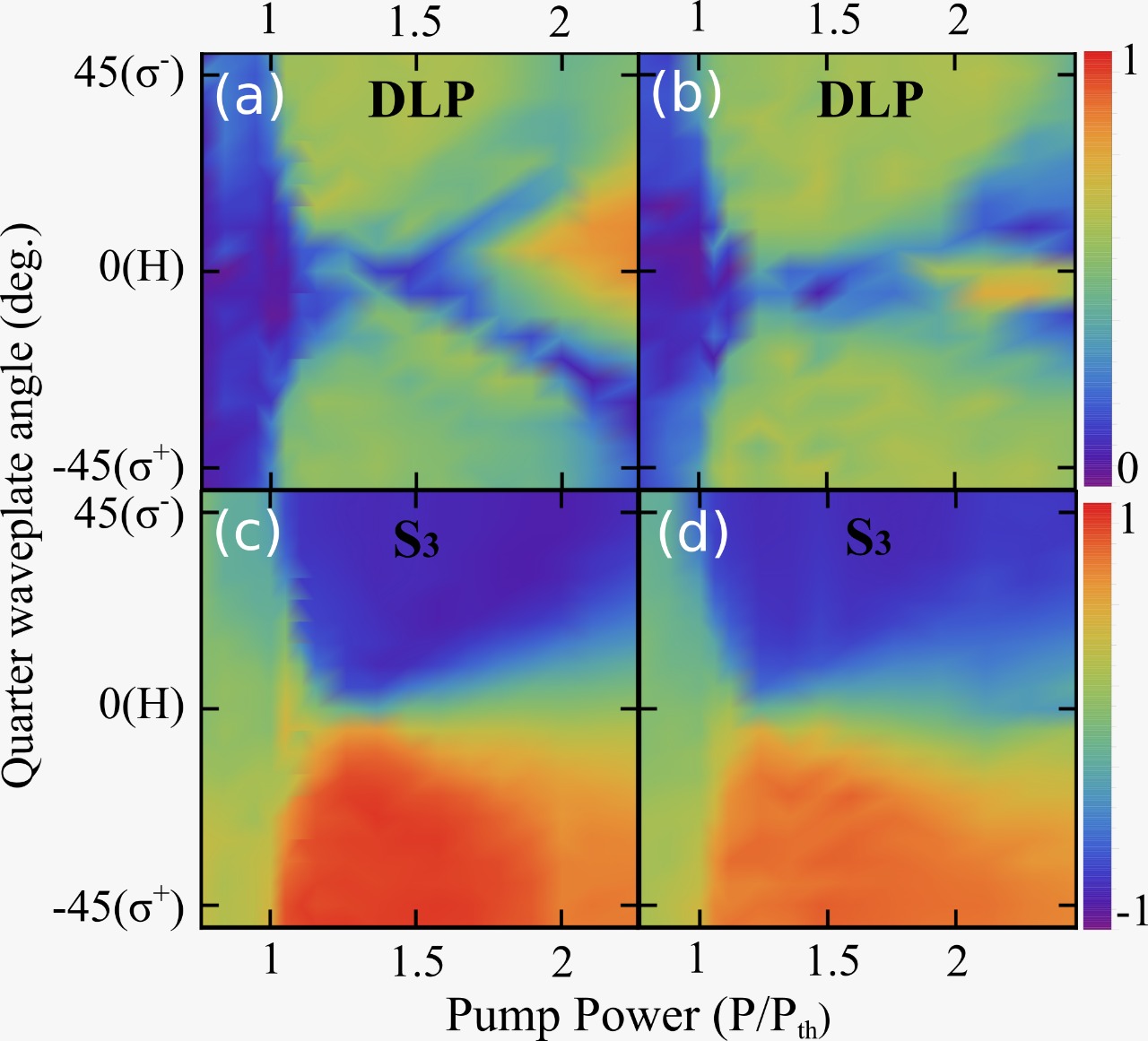}
    \caption{Measured DLP (a,b) and $S_3$ (c,d) for a detuning of $\Delta = -4$ meV (a,c) and $\Delta = -2$ meV (b,d). }
    \label{detuning}
\end{figure}

Here we present results on the condensate polarization map for two more detuning values $\Delta = -4$ meV and $\Delta = -2$ meV for the ring of diameter $d=12$ $\mu$m (see Fig.~\ref{detuning}). We point out that the detuning corresponding to the results of the main text is around $\Delta \approx -3$ meV.

The overall trend for these two additional measurements is the same as shown in the main text. We observe optical orientation under circularly polarized pumps (i.e., the condensate circular polarization follows that of the pump), and formation of the linear polarization island at big excitation power, accompanied with depolarized streaks [see Fig.~\ref{detuning}(a)]. The linearly polarized island appears around the same range of the excitation powers, but the shape of it is a bit different, which could be caused by disorder in the sample.

\section{Sample birefrigence measurement}
The microcavity that we studied, like most microcavity samples, has an inherent birefringence which is also anisotropic across the sample. To measure the birefringence of our sample, we use a set of two polarizers and illuminate the sample with a variable linear polarization of our excitation laser at normal incidence and without any focusing optics to define the fast and slow axes of the sample. In order to extract the retardance we operate in the basis of the sample's fast and slow axis. Utilizing the Muller matrix representation for an arbitrary retarder with the slow axis at zero degrees, we illuminate our sample with diagonal polarization in order to find the sample retardance $\delta$.
\begin{equation}
\begin{bmatrix}
1 & 0 & 0 & 0 \\
0 & 1 & 0 & 0 \\
0 & 0 & \cos{(\delta)} &-\sin{(\delta)} \\
0 & 0 & \sin{(\delta)} & \cos{(\delta)} 
\end{bmatrix} 
\begin{bmatrix}
1 \\ 0 \\ 1 \\ 0 
\end{bmatrix}  
=
\begin{bmatrix}
1 \\0 \\ \cos{(\delta)} \\ \sin{(\delta)}
\end{bmatrix}.
\end{equation}
For the experimental excitation wavelength ($\lambda_{exc}=\unit[783.6]{nm}$), we find a $\delta_{exc} \approx 0.07 \pi$. This small birefringence can explain the small rise of the $S_3$ component just above condensation threshold for diagonal and anti-diagonal polarized pumps, as shown in Fig.~5(a) of the main text. For linear polarization angles of the excitation that do not coincide with the fast or slow axes of the sample, the light gains a small degree of polarization ellipticity while propagating through the strained cavity mirrors before exciting electrons and holes in the cavity. 

Finally, we examine the sample birefringence for the condensate emission wavelength ($\lambda_{em}=\unit[856.6]{nm}$). For this wavelength we find a retardance of $\delta \approx 0.15 \pi$ and $\delta \approx 0.06 \pi$ in transmission and reflection respectively.

\section{Numerical Simulations and theoretical description}
\subsection{Full Stokes map characterization}
\begin{figure}
    \centering
    \includegraphics[width=1\columnwidth]{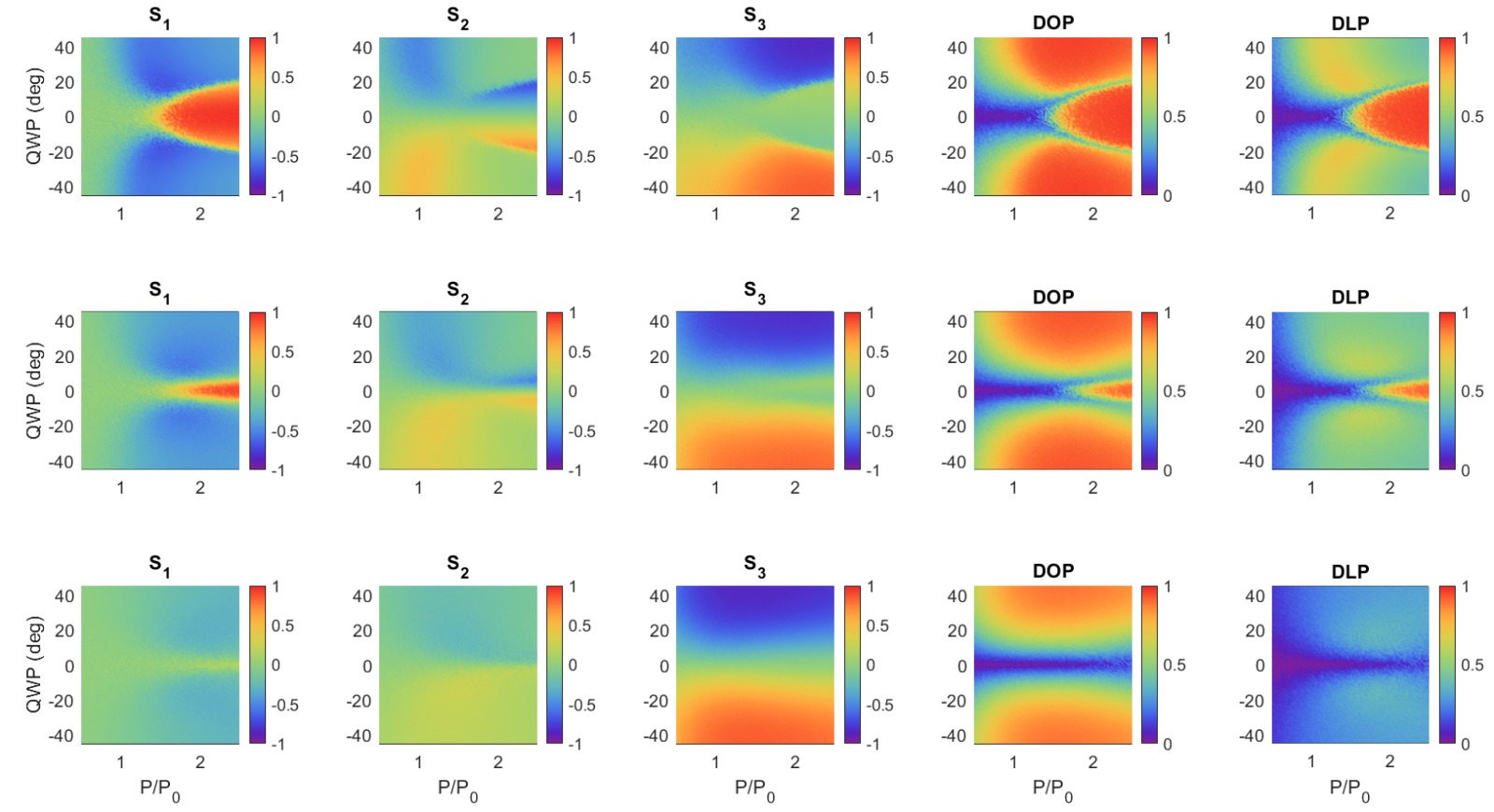}
    \caption{Average Stokes components $\bar{S}_{1,2,3}$, DOP, and DLP plotted as a function of QWP angle and pump power. The integration (simulation) time window is $T = 10$ ns. Pumping power is given in units of $P_0 =2 \Gamma\Gamma_A(\Gamma_I+W) / RW$. }
    \label{figfullmap}
\end{figure}

In Fig.~\ref{figfullmap} we present the rest of the simulated average Stokes components $\bar{S}_{1,2,3}$ as well the DOP and DLP corresponding to the simulation in Fig.~4 in the main text. 

\subsection{Condensate pinning as a function of power}
\begin{figure}
    \centering
    \includegraphics[width=0.8\columnwidth]{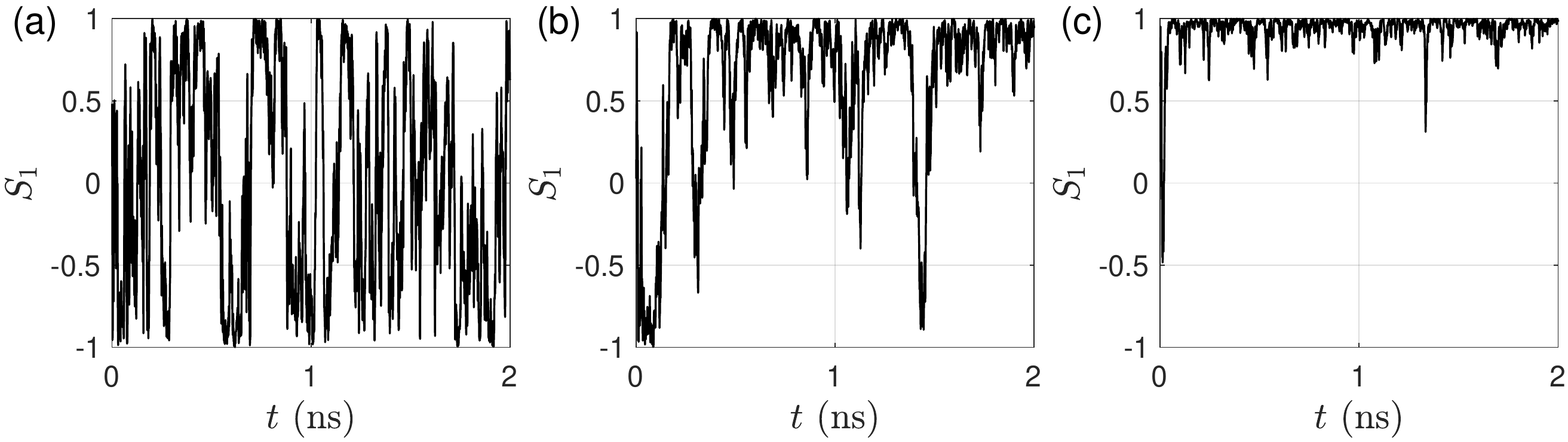}
    \caption{Simulated $S_1$ Stokes component for QWP $=0^\circ$ showing the onset of pinning with growing pump power (nonlinearity). (a) $P = 1.2P_0$. (b) $P = 1.5P_0$. (c) $P = 2 P_0$. }
    \label{fig.pin}
\end{figure}

In Fig.~\ref{fig.pin} we show the simulated onset of pinning at QWP $=0^\circ$ for increasing pump power. Panels (a-c) correspond to pump power $P = (1.2; 1.5; 2) P_0$ with $P_0$ defined in the main text and the caption of Fig.~\ref{figfullmap}. The results show that the condensate pseudospin stabilizes when the condensate nonlinearity, which scales with pump power, increases.

\subsection{Destabilization of the pinned polarization island}
In this subsection, we simplify our equations of motion [Eqs.~(1)-(3) in main text] and adopt a reservoir free model [Ohadi et al., Phys. Rev. X \textbf{5}, 031002 (2015)] to describe the stability properties of the polarization island. Such a model becomes accurate when the characteristic timescales of the reservoirs are shorter than that of the condensate.
\begin{equation}
\partial_t \psi_\sigma =\theta_\sigma(t) +  \left[ (1 - ig)P_\sigma  - (R + i \alpha) |\psi_\sigma|^2 \right]\psi_\sigma + i \Omega_x \psi_{-\sigma}.
\end{equation}
Here, $P_\sigma$ denotes the net injection rate of particles through the polarized nonresonant excitation, $R$ saturates the condensate above threshold, $gP_\sigma$ is blueshift originating from interactions with an excitonic reservoir scaling with the pump excitation, $\alpha$ is the polariton-polariton interaction strength, $\Omega_x$ is the in-plane magnetic field which couples the spin components of the condensate, and $\theta_\sigma(t)$ is a white noise term. 

\begin{figure}
    \centering
    \includegraphics[width=0.8\columnwidth]{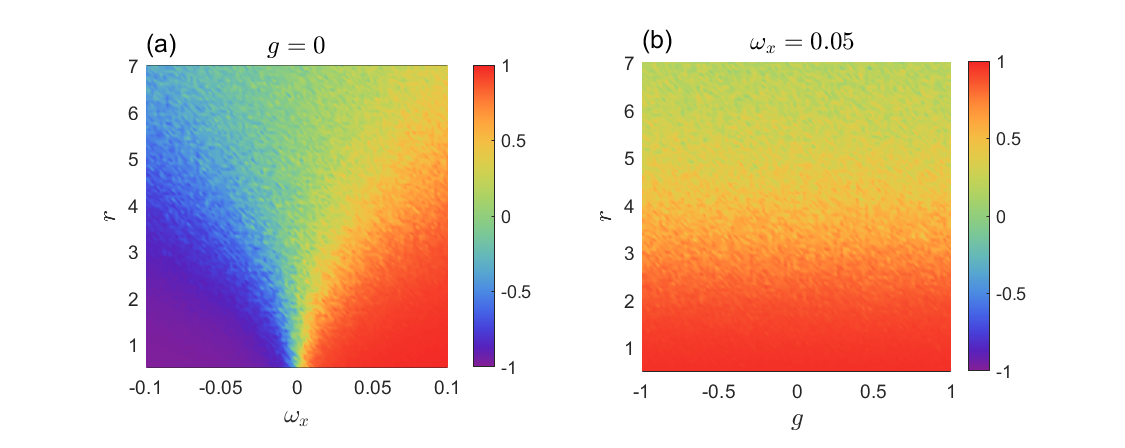}
    \caption{Simulated $\bar{S}_1$ Stokes component using Eq.~\eqref{eq.GP}. (a) Varying the coupling $\omega_x$ changes both the sign an strength of the polarization island. As $r$ increases the amplitude of the wavefunction (i.e., the polariton population $\bar{S}_0 = 2/r$) decreases and the polarization smears out against fluctuations going to zero. (b) Here $\omega_x = 0.05$ and $g$ is varied showing that for linearly polarized excitation the blueshift coming from the pump does not change the average degree of condensate linear polarization against fluctuation.}
    \label{fig.pin2}
\end{figure}

Let us consider the case where $P_\sigma = P$ which corresponds to linearly polarized excitation, QWP $=0^\circ$. We additionally rescale the wavefunction $\psi_\sigma = \psi'_\sigma\sqrt{P/\alpha}$ and define time in units of pump power, $t= \tau/P$ for brevity.
\begin{equation}\label{eq.GP}
\partial_\tau \psi'_\sigma = \theta_\sigma(t) + \left[ 1 - i g - (r + i) |\psi'_\sigma|^2 \right]\psi'_\sigma + i \omega_x \psi'_{-\sigma}
\end{equation}
where $r = R/P$, and $\omega_x = \Omega_x/P$. In Fig.~\ref{fig.pin2} we show the $\bar{S}_1$ Stokes component for two slices of the three dimensional parameter space of Eq.~\eqref{eq.GP}. In Fig.~\ref{fig.pin2}(a) we see that as $r$ grows the averaged steady state polariton occupation $\bar{S}_0 = 2/r$ decreases and the $\bar{S}_1$ components smears out against fluctuations and goes to zero. As $|\omega_x|$ becomes stronger the polaritons are more easily pinned even for small occupation number of the condensate. In Fig.~\ref{fig.pin2}(b) the coupling is fixed at $\omega_x = 0.05$ and we vary now both $g$ and $R$. The results show that the diagonal shift in energy does not affect the stability of the polarization island against random fluctuations. This is an expected result because we can always choose a rotating reference frame for the polaritons which rotates at frequency $g$. Thus, the formation of the polarization island as seen in experiment at QWP $=0^\circ$ can be attributed purely to the balance between polariton nonlinearity and the in-plane effective field that couples the two spin components.

\end{document}